# Similarity Downselection: A Python implementation of a heuristic search algorithm for finding the set of the *n* most dissimilar items with an application in conformer sampling

Felicity F. Nielson, Sean M. Colby, Ryan S. Renslow*, Thomas O. Metz*
Pacific Northwest National Laboratory, Richland, WA, USA
*Corresponding authors: ryan.renslow@pnnl.gov, thomas.metz@pnnl.gov

**Abstract**
Finding the set of the *n* items most dissimilar from each other out of a larger population becomes increasingly difficult and computationally expensive as either *n* or the population size grows large. Finding the *set* of the *n* most dissimilar items is different than simply sorting an array of numbers because there exists a pairwise relationship between each item and all other items in the population. For instance, if you have a set of the most dissimilar *n=4* items, one or more of the items from *n=4* might not be in the set *n=5*. An exact solution would have to search all possible combinations of size *n* in the population, exhaustively. We present an open-source software called similarity downselection (SDS), written in Python and freely available on GitHub. SDS implements a heuristic algorithm for quickly finding the approximate set(s) of the *n* most dissimilar items. We benchmark SDS against a Monte Carlo method, which attempts to find the exact solution through repeated random sampling. We show that for SDS to find the set of *n* most dissimilar conformers, our method is not only orders of magnitude faster, but is also more accurate than running the Monte Carlo for 1,000,000 iterations, each searching for set sizes *n=3–7* out of a population of 50,000. We also benchmark SDS against the exact solution for example small populations, showing SDS produces a solution close to the exact solution in these instances.

## 1. Introduction

There exists a large and historical body of algorithms and their implementations for searching, sorting, and clustering data based on distance or similarity. Popular algorithms include beam search and K-means clustering [1-4], used for finding the target result by following the most promising nodes (as determined by an evaluation function *f(n)*) and for grouping data by similarity to a number of selected data points or nodes, respectively. One similarity-comparison problem involves choosing the set of the *n* most dissimilar items from a larger population of size *N*, in which there exists a pairwise relationship between each item and all other items. There exists several older algorithms in the literature to solve this type of problem [5-7], but none are available as open-source Python packages. The solution to this dissimilarity-set problem is useful in chemistry and biology, for instance, for finding the most geometrically dissimilar sets of conformers (or molecular structures) to efficiently span conformational space and eliminate redundant structures. The use of root-mean-square deviation (RMSD) of atomic positions for selecting sets of conformers has been used by many groups [8-12].

Finding the exact solution to the most dissimilar set problem becomes intractably computationally expensive (super-exponentially complex) as the population size *N* grows large and is most expensive when *n=N/2*, according to the binomial coefficient

$$\binom{N}{n} = \frac{N!}{(N-n)!\,n!} \quad (1)$$

Additionally, if the set *n=4* is found, one or more of its items might not be a member of the set *n=5*. Verifying the exact solution has been found requires exhaustively searching all possible combinations for a given set size. Finding the exact solution quickly becomes intractable for classical computers. When considering a population of 50,000 items, for example, finding the most dissimilar set of *n=3* items would require searching over 20 trillion unique combinations (2.083208335e13 to be exact). However, if one assumes all members of the set *n* are also members of the set *n+1*, then an approximate solution that is sufficiently close to the exact solution can be determined in a "greedy" fashion. We introduce a heuristic algorithm implemented in Python, similarity downselection (SDS), that finds the subset of the *n* most dissimilar items from a large population. SDS is generalizable to any application where the data can be represented as arrays whose elements are the pairwise relationships between each item and all other items in the population. We include a brief description of an example application on molecular conformer selection, and benchmark SDS against both a Monte Carlo sampling method and the exact solution.

## 2. Similarity Downselection Python Module

SDS implements a heuristic algorithm for finding the set of $n$ items most dissimilar from each other out of a larger population. The algorithm is greedy, making the optimal choice during each iteration, where each iteration finds the set $n+1$ by building off the set $n$. SDS is freely available as a Python module on GitHub at https://github.com/pnnl/sds. Below, we provide short descriptions of how the algorithm works using arrays, and alternatively, using node/graph theory.

### 2.1 Algorithm Description

The individual items in a population are represented as arrays whose elements contain floating point values of the pairwise relation (e.g. RMSD or other dissimilarity metric) between the given item and all other items in the population. The first element of all arrays is reserved for the pairwise relation to the first item, the second element to the second item, and so forth until an $NxN$ matrix is formed, where $N$ is the size of the total population, the $i^{th}$ row is the array of the $i^{th}$ item, and $N_{ij}$ contains the pairwise relation between items $i$ and $j$. Since $N_{ij} = N_{ji}$, the matrix is symmetric across the diagonal.

The algorithm first selects the two items that have the largest pairwise value between them. This is the exact subset of $n=2$ most dissimilar items. To find the subset $n=3$, the natural log of the first two arrays (i.e. the first two rows corresponding to the first two items), are summed element wise to create a new summation array. The index of the largest value in the summation array is the index of the third most dissimilar item. The natural log of the array corresponding to the third item is then added to the summation array to yield the index of the fourth most dissimilar item. Successive subsets $n=5, 6, 7, …, N,$ are achieved in the same manner, selecting the item corresponding to the index of the highest value.

Items could be multiplied into a multiplication array instead of log-summed, but it was found the product of floating point numbers quickly exceed machine precision (10e-323 in our setup) after about 10,000 items, so log-summing is used instead. Effectively, log-summing (or multiplying) rewards items that have a large value across all arrays by making its numerical representation larger and punishes items that have even one significantly small pair-wise relation with another item by making its numerical representation smaller. $N$ can be very large, theoretically indefinite and limited only by machine precision and memory. The population used for the original implementation, as discussed in Section 3, contained 50,000 items.

### 2.2 Problem and algorithm description using graph theory (nodes and edges)

Like the traveling salesman problem, the longest path problem (LPP), and many other problems found in graph theory, the items in the most dissimilar set problem (MDSP) can be thought of as nodes. Like LPP for weighted complete graphs (where each node is connected to every other node, and the edges are assigned weights), MDSP seeks to find the nodes that will maximize the total distance. More specifically, MDSP must find the subset of size $n$ that will yield a maximum distance. In graph theory, this makes MDSP more general than LPP because LPP is a special case when $n=N$. In MDSP permutation, unlike LPP, the order the nodes are visited does not matter. For example, when the target set size is the full population size ($n=N$), the solution to MDSP is trivially the full population. In contrast, the solution for LPP has not only not been found, it has not even been searched for. Additionally, while problems like LPP require the total distance be calculated by the simple path traveled between nodes (only two connecting edges per node, entering and exiting), MDSP takes into account all pairwise edges in the solution set exactly once. In other words, the exact solution to MDSP can and must travel all pairwise edges in the chosen set and does not double count edges that have already been traveled.

SDS uses a one-dimensional representation to inform its decision traversing the graph. The algorithm starts by finding the two nodes most distant from each other (two nodes with the highest weight assigned to their connecting edge). It then takes the one-dimensional representation of one of the two nodes and log-sums its weighted edges (distances) with the one-dimensional representation of the second node. This creates a new one-dimensional representation where all previously chosen nodes have zero distance and the node with the furthest distance (highest log-sum) is the next node chosen on the graph.

## 3. Example Application: Molecular Conformer Sampling

A method was needed to find the subset of the $n$ most dissimilar molecular conformers (instances of the same molecule containing the same atoms and bonds but with different geometrical structures) out of a set of 50,000 conformers, where $1 < n < 50000$. In this application, the goal was to efficiently span conformer space by downselecting from a larger population to the most structurally dissimilar conformers, as part of a larger analysis to assess the validity of various conformer selection techniques [13]. The dissimilarity between two conformers was measured as the average pairwise RMSD between corresponding atoms, calculated using OpenBabel (v 2.4.1) [14, 15]. The method for finding the *nth* dissimilar set needed to be efficient and applicable to any small molecule.

## 4. Benchmarking
### 4.1 Performance against a Monte Carlo method
SDS was shown to be faster and produce more dissimilar sets than a Monte Carlo (MC) sampling method in a contest to find the most dissimilar sets *n=3–7* out of a population of 50,000 conformers for sphingosine $[M+H]^+$. MC sampling was run for 1,000,000 iterations for each *n*-sized set, taking more than 2 hours to complete for each. After loading the data matrix, which required about 3 min, the heuristic algorithm found all sets in <1 min. SDS also had a greater RMSD log-sum (total distance between nodes) for every set size, as shown in **Fig. 1**, indicating it was closer to the exact solution than the MC method every time.

This benchmarking analysis was applied again to 50,000 conformers of methyleugenol $[M+Na]^+$ with similar results. Here, MC performed better than SDS at *n=3* by a small margin (). SDS ran the complete search for every possible set *1 < n < 50000* in approximately 7 min, including the approximate 3 min required to load the matrix.

### 4.2 Performance against the exact solution
SDS was benchmarked against the exact solution for *N=20, 22, 24* with *n=N/2* on randomly generated datasets, as summarized in **Fig. 2**. In each case, the SDS solution had a total distance closer to the exact solution distance than the mean set, indicating a good heuristic solution.

In our setup, we estimate it would require over 72 node hours to find the exact solution for a population even as small as *N=30* with *n=15*, while SDS would find a heuristic solution in a fraction of a second.

### 4.3 Comparing computational cost of calculating pairwise relations
SDS requires the pairwise relation matrix to be calculated in advance. Depending on the application and what set size is being searched for, initially this may falsely appear to

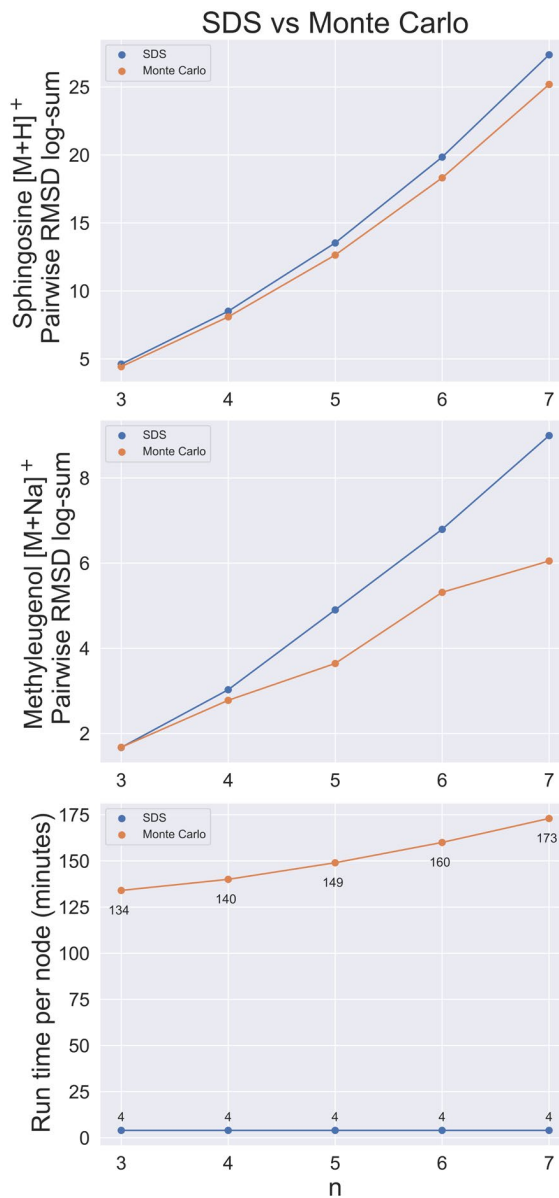

Fig. 1 SDS benchmarked against a Monte Carlo (MC) sampling method for sphingosine $[M+H]^+$ and methyleugenol $[M+Na]^+$ with conformer populations of 50,000. **Top and Middle**, the conformer RMSD log-sum (a metric of the dissimilarity of the set) for SDS and the largest RMSD log-sum found by MC at set size *n*. **Bottom**, search time per node for both methods. Time includes the (approximate) 3 minutes to load the pairwise RMSD matrix.

reduce the cost-effectiveness. Because the same pairwise relations would have to be calculated for both the MC method and the exhaustive search, the total computational cost is equal to the cost for SDS, assuming each relation is efficiently calculated only once during each method. If MC fails to consider every possible pairwise relation and is computing them on the fly, then there would be fewer pairwise relations to compute, but this would be the same as creating a randomly selected subset of data and running

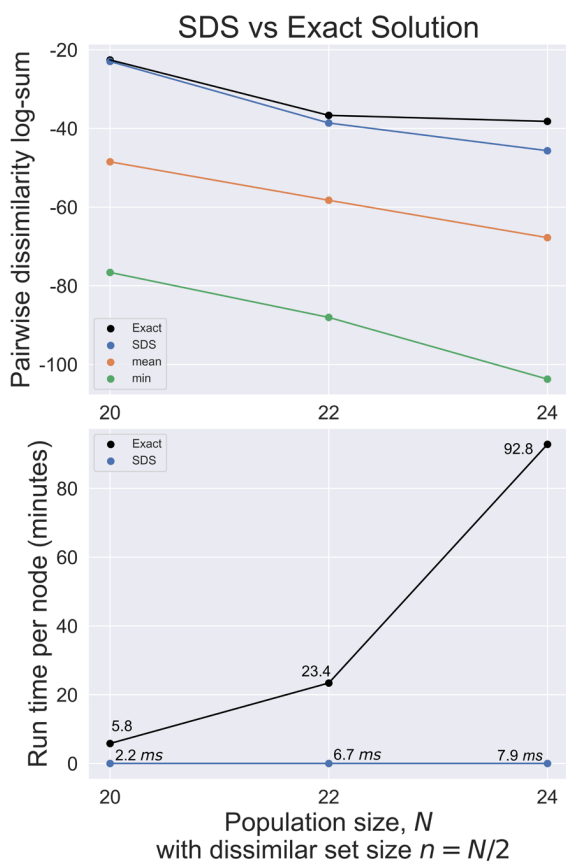

Fig. 2 SDS benchmarked against the exact solution on randomly generated datasets with population size *N*, searching for the most dissimilar set of size *n=N/2*. **Top**, total pairwise dissimilarity for the exact solution, SDS, mean, and minimum (most similar) sets. **Bottom**, search time per node for both methods.

MC searches when SDS could have been run on the subset just as well, maintaining the benefit of SDS.

## 5. Conclusion

We have introduced a new software written in Python implementing a heuristic algorithm for finding the set of *n* items most dissimilar from each other and demonstrated its efficacy and efficiency in benchmarks against a Monte Carlo method and the exact solution. SDS, freely available at https://github.com/pnnl/sds, has application in molecular conformer selection, but also has potential application in searches for the $n^{th}$ most dissimilar set in generalized datasets.

## 6. Acknowledgments

This work was supported by the National Institutes of Health, National Institute of Environmental Health Sciences grant U2CES030170. Pacific Northwest National Laboratory (PNNL) is operated for the U.S. Department of Energy by Battelle Memorial Institute under contract DE-AC05-76RL01830.